\documentclass[aps,prl,reprint,groupedaddress]{revtex4-2}

\usepackage{graphicx}
\usepackage{graphics}
\usepackage{epsfig}
\usepackage{epstopdf} 
\usepackage{amsmath}
\usepackage{caption}
\usepackage{xcolor}

\begin{document}

\title{Precise measurement of angles between two magnetic moments and their configurational stability in single-molecule magnets}

\date{\today}

\author{Rasmus Westerstr\"om}
\email{rasmus.westerstrom@sljus.lu.se}
\affiliation{Synchrotron Radiation Research, Lund University, SE-22100 Lund, Sweden}
\affiliation{NanoLund, Lund University, SE-22100 Lund, Sweden}
\author{Vasilii Dubrovin}
\affiliation{Leibniz Institute of Solid State and Materials Research, D-01069 Dresden, Germany}
\author{Katrin Junghans}
\affiliation{Leibniz Institute of Solid State and Materials Research, D-01069 Dresden, Germany}
\author{Aram Kostanyan}
\affiliation{Physik-Institut, Universit\"at Z\"urich, CH-8057 Z\"urich, Switzerland}
\affiliation{Swiss Light Source, Paul Scherrer Institut, CH-5232 Villigen PSI, Switzerland}
\author{Christin Schlesier }
\affiliation{Leibniz Institute of Solid State and Materials Research, D-01069 Dresden, Germany}
\author{Jan Dreiser}
\affiliation{Swiss Light Source, Paul Scherrer Institut, CH-5232 Villigen PSI, Switzerland}
\author{Bernd B\"uchner}
\affiliation{Leibniz Institute of Solid State and Materials Research, D-01069 Dresden, Germany}
\author{Stanislav M. Avdoshenko}
\affiliation{Leibniz Institute of Solid State and Materials Research, D-01069 Dresden, Germany}
\author{Alexey A. Popov}
\affiliation{Leibniz Institute of Solid State and Materials Research, D-01069 Dresden, Germany}
\author{Thomas Greber}
\affiliation{Physik-Institut, Universit\"at Z\"urich, CH-8057 Z\"urich, Switzerland}

\date{\today}

\begin{abstract}
A key parameter for the low-temperature magnetic coupling of in dinuclear lanthanide single-molecule magnets (SMMs) is the barrier $U_{FA}$ resulting from the exchange and dipole interactions between the two $4f$ moments. Here we extend the pseudospin model previously used to describe the ground state of dinuclear endofullerenes to account for variations in the orientation of the single-ion anisotropy axes and apply it to the two SMMs Dy$_2$ScN@C$_{80}$ and Dy$_2$TiC@C$_{80}$.
While x-ray magnetic circular dichroism (XMCD) indicates the same $J_z=15/2$ Dy groundstate in both molecules, the Dy-Dy coupling strength and the stability of magnetization is distinct. We demonstrate that both the magnitude of the barrier  $U_{FA}$ and the angle between the two  $4f$ moments are determined directly from precise temperature-dependent magnetization data to an accuracy better than $1^{\circ}$. The experimentally found angles between the $4f$ moments are in excellent agreement with calculated angles between the quantisation axes of the two Dy ions. Theory indicates a larger deviation of the orientation of the Dy magnetic moments from the Dy bond axes to the central ion in Dy$_2$TiC@C$_{80}$. This may explain the lower stability of the magnetisation in Dy$_2$TiC@C$_{80}$, although it exhibits a $\sim 49\%$ stronger exchange coupling than in Dy$_2$ScN@C$_{80}$.  
\end{abstract}

\maketitle

%

Stabilizing magnetic moments of single atoms is an active research field motivated by applications in molecular spintronics, quantum computation, and the quest for the ultimate miniaturization of data storage \cite{Bogani_matNat2011,Leuenberger_nat2001}. The required anisotropy for stabilizing magnetic moments or spins is realized in single-molecule magnets (SMMs). They exhibit hysteresis below a certain blocking temperature at which the relaxation of the magnetization becomes slow compared to the measurement time  \cite{Gatteschi_Molecular_2006,Sessoli_Magnetic_1993,Benelli_Introduction_2018,Ishikawa_jacs2003,HARRIMAN2019425,C8DT05153D,C7CS00266A}.

\begin{figure}[t]
\begin{center}
\includegraphics[width=8.5cm]{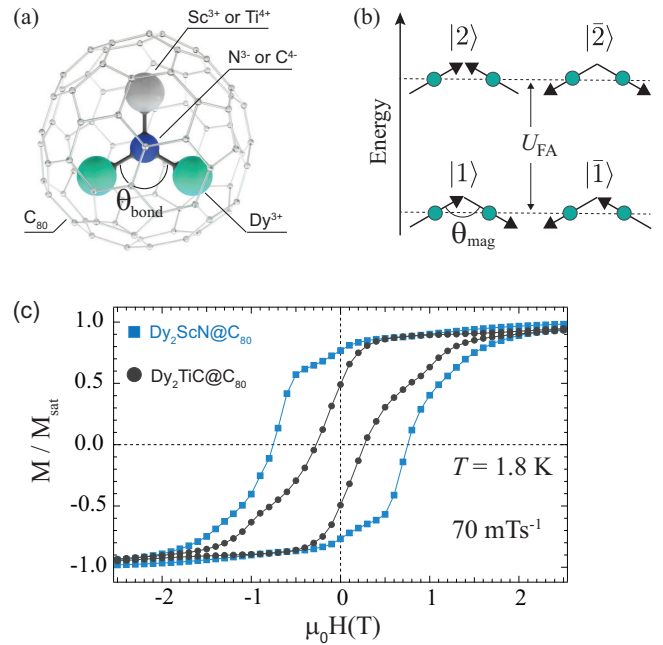}
	\caption{(a) Ball and stick-model of Dy$_2$ScN@C$_{80}$ and Dy$_2$TiC@C$_{80}$. (b) Pseudospin ground state model in the absence of an applied magnetic field. The arrows indicate the orientation of the magnetic moments along the anisotropy axis on each Dy site. The mutual orientation of the anisotropy axis in the $\mathrm{Dy}-\mathrm{Dy}$ dimer is given by the angle $\theta_{mag}$. Exchange and dipole interactions lead to a  ground state with two degenerate time-reversal symmetric and ferromagnetically coupled doublets  $\vert 1 \rangle$ and $\vert \bar{1}\rangle$.  Quantum tunneling of the magnetization is blocked in the ground state, and the relaxation proceeds via the antiferromagnetic doublets $\vert 2 \rangle$ and $\vert \bar{2}\rangle$. Reversing the magnetization is thus associated with an exchange and dipole barrier $U_{FA}$, corresponding to the energy gap between the two doublets. (c) Magnetization curves recorded at 1.8 K using SQUID magnetometry at a field sweep rate of 70 mTs$^{-1}$. Element-specific magnetization curves recorded at the Dy $M_5$-edge are shown in the supplementary \cite{SI} with a strong resemblance to those in (c), while being recorded at slightly higher temperature and exposed to X-rays \cite{deMag}.}
\label{fig1}.
\end{center}
\end{figure}

Magnetic bistability of a single ion was first demonstrated for double-decker TbPc$_2$ lanthanide complexes \cite{Ishikawa_jacs2003} and later for endofullerenes exhibiting longer magnetic lifetimes \cite{westerstromJACS}.  The interaction between the $4f$ orbital and the ligand field (LF) creates an anisotropy barrier separating states of different magnetization and thereby provides a prerequisite for stabilizing single magnetic moments.  However, the presence of a large anisotropy barrier is not sufficient for a stable remanent magnetization due to the possibility of shortcutting the anisotropy barrier by quantum tunneling of magnetization (QTM). QTM is influenced by hyperfine interaction or dipolar stray fields from neighboring SMMs \cite{westerstromJACS}, coupling to the phonon bath \cite{Dy2LuN} and off axis components of the $g$-tensor \cite{chibotaru}. In lanthanide single-ion magnets, QTM is observed in the hysteresis as a sharp drop in the magnetization close to zero field, which drastically reduces the remanent magnetization \cite{westerstromJACS,westerstromPrb14}. Suppression of QTM can be achieved by minimizing intermolecular interactions through dilution \cite{krylovDy1}, adsorbing the molecule on a suitable substrate \cite{Wckerlin_Single_2016} or coupling of two or more lanthanide ions in polynuclear complexes \cite{westerstromPrb14,popov2017,Rinehart_A_2011,Rinehart_Strong_2011,Velkos_High_2019,Guo2011}. For  Dy$_2$ScN@C$_{80}$, the latter results in a ferromagnetically (FM) coupled ground state where relaxation proceeds via the antiferromagnetically (AFM) coupled states, thereby stabilizing the remanent magnetization at low temperatures with a protection barrier $U_{FA}$ \cite{westerstromPrb14}. A  significant remanent magnetization is also observed at 1.8 K for the isoelectronic and isospintronic didysprosium sister compound Dy$_2$TiC@C$_{80}$.  However, Dy$_2$TiC@C$_{80}$ is  magnetically less stable compared to Dy$_2$ScN@C$_{80}$  \cite{Junghans}, as evident from the hysteresis in Fig. \ref{fig1} (c) where Dy$_2$TiC@C$_{80}$ has a lower remanent magnetization and a significantly smaller coercive field.  Research aims to understand such differences, which will pay off for the design of better SMMs. Here we explore with magnetisation measurements the limits of experimental determination of angles between magnetic moments, which are key quantities in molecular magnetism.

The present paper applies the pseudospin model in  Fig.\ref{fig1} to temperature-dependent magnetization data from Dy$_2$ScN@C$_{80}-I_h$ and Dy$_2$TiC@C$_{80}-I_h$ (hereafter the isomeric label $I_h$ is omitted for clarity) to extract values of the protection barrier $U_{FA}$. We find a \textit{higher} barrier for Dy$_2$TiC@C$_{80}$ even though the zero-field lifetimes are much \textit{shorter},  demonstrating that a stronger coupling not necessarily provides greater resistance towards demagnetization. Careful evaluation of the magnetization data further provides angles between the two magnetic moments in the Dy$_2$ dimers with an accuracy better than 1 degree. Since the angle between the moments enters all expressions for the description of the mutual $\mathrm{Dy} - \mathrm{Dy}$ interaction, our results open new perspectives for both, testing and improving theories and magnetic materials. As a first demonstration of the discriminative power of accurate angle determination, we show that the experimentally determined angles fit the $\mathrm{Dy}-\mathrm{X}-\mathrm{Dy}$ $(\mathrm{X=N, C})$ bond angles from density functional theory (DFT) worse than the angles between the two quantization axes of the $\mathrm{Dy}$ magnetic moments from complete active space self-consistent-field (CASSCF) calculations. This satisfies the expectation and increases confidence in experiment and theory. Notably, the larger off-axis $g$-tensor components of  Dy$_2$TiC@C$_{80}$  are in line with the shorter zero-field lifetimes. The larger off-axis $g$-tensor components of Dy$_2$TiC@C$_{80}$ indicate, compared to Dy$_2$ScN@C$_{80}$, a lower axial symmetry, and a larger rhombicity \cite{Liu2014}. This stronger mixing of different $J_z$ levels may decrease zero-field lifetimes. Therefore, the angle between the magnetic moment and the molecular bond axis is an essential quantity for understanding the stability of the magnetisation in SMMs.

%

%
\section{EXPERIMENTAL} 

The Dy$_2$ScN@C$_{80}$ (Dy$_2$TiC@C$_{80}$) endofullerenes were produced using an arc-discharge synthesis using graphite rods packed with a mixture of Sc (Ti), Dy, and graphite powder under He atmosphere with small amounts of NH$_3$ (CH$_4$)\cite{westerstromPrb14,Junghans}. The X-ray absorption measurements were carried out at the X-Treme beamline \cite{piamontezeXtreme} of the Swiss Light Source. Absorption spectra were acquired by measuring the total electron yield (TEY) in the on-the-fly mode \cite{krempasky} while applying a magnetic field parallel to the X-ray beam. The SQUID measurements were performed using a Quantum Design MPMS3 Vibrating Sample Magnetometer (VSM). The endofullerenes were dissolved in toluene and spray-coated on an aluminum plate and drop-cast into a polypropylene sample holder for the XMCD and SQUID measurements respectively. 
\section{THEORY} 
DFT calculations for isolated Dy$_2$ScN@C$_{80}$ and Dy$_2$TiC@C$_{80}$ molecules were performed at the PBE-D level with a plane-wave basis set and corresponding projector augmented-wave potentials, treating $4f$-electrons as a part of the core as implemented in the VASP 5.4 package  \cite{Hafner,Kresse1993,Kresse1999,Perdew1996,Grimme}. Ab initio calculations of the multiplet structure and pseudospin $g$-tensors for the DFT-optimized conformers of Dy$_2$ScN@C$_{80}$ and Dy$_2$TiC@C$_{80}$ were performed at the CASSCF(9,7)/SO-RASSI level using the quantum chemistry package OpenMOLCAS \cite{Molcas} and its SINGLE\_ANISO module \cite{Chibotaru2012} VDZ-quality atomic natural extended relativistic basis set (ANO-RCC) was employed for inner clusters and ANO-RCC-MB for carbon cages. In each calculation, one Dy atom was treated \textit{ab initio} and another one was replaced with Y.

\section{RESULTS and DISCUSSIONS} 

%

Earlier, we found that the single-ion magnets DySc$_2$N@C$_{80}$ and DyYTiC@C$_{80}$ exhibits comparable hysteresis closing temperatures and relaxation times, indicating that the single-ion anisotropy has a similar influence on the magnetic bi-stability for the Dy-nitride and Dy-carbide cluster fullerenes \cite{C8CC04736G}.  In these two systems, the central non-metal ion, N$^{3-}$ or C$^{4-}$, provides a LF that lifts the degeneracy of the $^6H_{15/2}$ Hund's ground state multiplets of the Dy$^{3+}$ ion and stabilizes an $J_z=15/2$ ground state with a quantization axis along the $\mathrm{Dy}-\mathrm{X}$ direction and nominal magnetic moment $\mu=10~\mu_B$ \cite{westerstromJACS,westerstromPrb14}. To compare the single-ion ground states in the di-dysprosium SMMs Dy$_2$ScN@C$_{80}$ and  Dy$_2$TiC@C$_{80}$,  we performed  XMCD measurements at the Dy $M_{4,5}$-edge, see \cite{SI}. The sum-rule \citep{Carra, Thole} results in Table \ref{tabel} from the two systems are within the accuracy of the experiment identical and in good agreement with previous studies of Dy$_2$ScN@C$_{80}$ \cite{westerstromJACS,westerstromNanoscale18}, confirming a  $J_z=15/2$ ground state in both compounds. Having established similar single-ion ground state properties, we turn to the $\mathrm{Dy}-\mathrm{Dy}$ interactions, which must be the root of the significantly different magnetic bistabilities observed in the two systems. 

%
\begin{table*}
\small
  \caption{Expectation values of the spin $\langle S_z\rangle$ and orbital $\langle L_z\rangle$ angular momentum operators and the resulting moment $\mu_z=-(\langle L_z\rangle+2\langle S_z\rangle)\mu_B$ resulting from a sum rule analysis of XMCD data measured at $\pm6.5$ T (see \cite{SI}). The single-ion moments $\mu$, and the exchange and dipole-barrier $U_{FA}$ was extracted from fitting the ground state model to the equilibrium magnetization curves in Fig. \ref{FigZ}. The angle $\theta_{mag}$ between the two Dy moments was obtained from fitting $\chi ^2(\theta)=\chi ^2_{\textrm{min}}+A(\theta-\theta_{mag})^2$ to the data in Fig.\ref{fig3},  whereas  $\Delta_{\mathrm{eff}}$ and  $\tau_0$ results from the Arrhenius plot in Fig. \ref{fig4}.  }
  \label{tabel}
  \begin{tabular*}{\textwidth}{@{\extracolsep{\fill}}lllllllll}
    \hline
    Sample & $\langle S_z \rangle~(\hbar)$ & $\langle L_z\rangle~(\hbar)$ &  $\langle \mu_z\rangle~(\hbar)$ & $\mu (\mu_B)$ & $U_{FA}(\mathrm{meV})$ &  $\theta_{mag}(^{\circ})$ &  $\Delta_{\mathrm{eff}}(\mathrm{meV})$ &  $\tau_0(\mathrm{s})$\\
    \hline
    Dy$_2$ScN@C$_{80}$  & $-1.4\pm0.1$ & $-2.3\pm0.1$ &  $5.1\pm0.1$ & $9.71\pm0.05$ & $0.71\pm0.01$ &  $117.1\pm0.8$ & $~~0.65$ & $70.8$\\
    Dy$_2$TiC@C$_{80}$  & $-1.4\pm0.1$ & $-2.3\pm0.2$ &  $5.1\pm0.2$ & $9.65\pm0.02$ & $1.06\pm0.01$ &  $115.2\pm0.7$ & $~~0.82$ & $5.1$ \\
    \hline
  \end{tabular*}
\end{table*}  

To compare the two systems, we adopt the pseudospin model previously applied to the ground state of the nitride cluster fullerene SMMs Dy$_2$ScN@C$_{80}$ \cite{westerstromPrb14}, Dy$_2$GdN@C$_{80}$ \cite{GreberGd}, and Tb$_2$ScN@C$_{80}$ \cite{GreberTb2}. Here, each moment can take two antiparallel directions along the magnetic easy-axis, resulting in  $2^2$ possible arrangements, grouped into two degenerate time-reversal symmetric  doublets, where the ground state corresponds to the FM configuration $\vert 1 \rangle$ and $\vert \bar{1}\rangle$, see Fig.\ref{fig1} (b). At 2-10 K, zero-field QTM between the ground-state doublets is blocked, and reversing the magnetization involves relaxation via the AFM doublets $\vert 2 \rangle$ and $\vert \bar{2}\rangle$ at an energy $U_{FA}$ \cite{westerstromPrb14}. Consequently, the magnetic relaxation at low temperatures becomes long compared to the measurement times,  resulting in hysteresis with large remanent magnetization and coercive field as shown in Fig.\ref{fig1} (c). However, for sufficiently slow field sweep rates, the system has time to reach thermal equilibrium,  and the magnetization curve will be reminiscent of a Brillouin function with a shape determined by the molecular moments in the FM and AFM doublets, the corresponding Zeeman interactions, and the barrier $U_{FA}$. The molecular moments are given by the vectorial sum of the single-ion moments $\mu$ in the FM and AFM configuration and consequently depend on the orientation of the corresponding anisotropy axes. To account for variations in the orientation of the single-ion anisotropy axes, we extend the model by introducing the angle $\theta_{mag}$ between the two magnetic moments, see Fig.\ref{fig1} (b). The equilibrium magnetization data thus allow us to extract information both about the magnetic $\mathrm{Dy}-\mathrm{Dy}$ interactions through $U_{FA}$, and the mutual orientation of the single-ion anisotropy axes. 

%
\begin{figure}[h]
\begin{center}
\includegraphics[width=8.5cm]{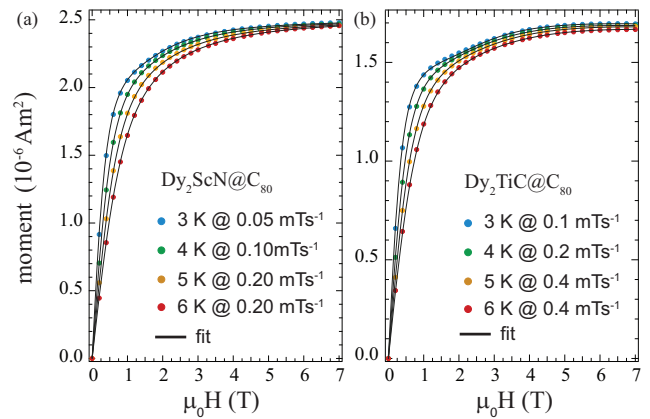}
	\caption{Equilibrium magnetization curves recorded for Dy$_2$ScN@C$_{80}$ and Dy$_2$TiC@C$_{80}$ using SQUID magnetometry (markers) at average field sweep rates between 0.05 mTs$^{-1}$ and 0.4 mTs$^{-1}$. The lines are the best fit of the ground state model to the data. }
\label{FigZ}.
\end{center}
\end{figure}

%
Figure \ref{FigZ} shows equilibrium magnetization curves recorded between 3 and 6 K at average field sweep rates ranging from  0.05 mTs$^{-1}$ to 0.4 mTs$^{-1}$ depending on the temperature. Fitting the ground state model to the magnetization curves yields single-ion magnetic moments of $9.71\pm0.05 ~\mu_B$ and $9.65\pm0.02~\mu_B$ for Dy$_2$ScN@C$_{80}$  and Dy$_2$TiC@C$_{80}$, respectively.  Surprisingly, the extracted barrier of $0.71\pm0.01$ meV for Dy$_2$ScN@C$_{80}$ is significantly lower than $1.06\pm0.01$ meV obtained for Dy$_2$TiC@C$_{80}$, indicating that a larger barrier does not necessarily lead to a more stable magnetic ground state. Similar to our experimental findings, broken-symmetry DFT calculations of Gd analogs predicted stronger FM coupling in Gd$_2$TiC@C$_{80}$ compared to Gd$_2$ScN@C$_{80}$ \cite{Chen2017}. However, it should be kept in mind that exchange coupling constants may not be transferable between Dy and Gd analogs.

\begin{figure}[h]
\begin{center}
\includegraphics[width=7cm]{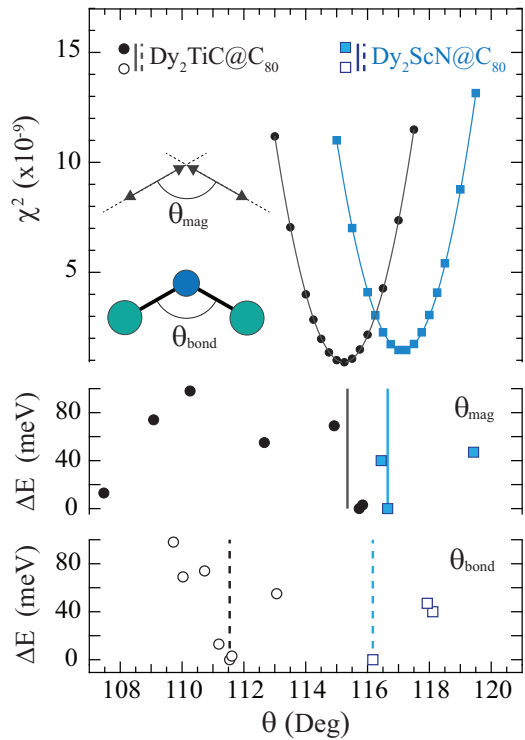}
	\caption{Top panel:  $\chi ^2$ curves from fitting the ground state model in Fig. \ref{fig1} to the magnetization curves in Fig. \ref{FigZ} for different angles $\theta$ between the single-ion anisotropy axes. Center panel: Angles between quantization axes $J_{z(1)}-J_{z(2)}$ of the two Dy moments in the Dy$_2$ScN@C$_{80}$ and Dy$_2$TiC@C$_{80}$ conformers. The vertical lines are the averaged values weighted with Boltzmann factors at a freezing temperature $T=60$ K for endohedral cluster rotation \cite{aramPrl}. Bottom panel: The corresponding $\mathrm{Dy}-\mathrm{X}-\mathrm{Dy}$ $(\mathrm{X=N, C})$ bond angles. 
}
\label{fig3}.
\end{center}
\end{figure}

As mentioned above, the magnetization curves contain information about the mutual orientation of the magnetic moments through the angle $\theta_{mag}$. To investigate the influence of the orientation of the anisotropy axis, we determined the $\chi ^2$ deviation between simulation and experiment for different angles $\theta_{mag}$. The resulting $\chi ^2(\theta_{mag})$ dependence in Fig. \ref{fig3} exhibits distinct minima from which $\theta_{mag}= 117.1\pm0.2^{\circ}$ for Dy$_2$ScN@C$_{80}$ and $\theta_{mag} = 115.2\pm 0.7^{\circ}$ for Dy$_2$TiC@C$_{80}$ are determined by fitting quadratic function $\chi ^2(\theta)=\chi ^2_{\textrm{min}}+A(\theta-\theta_0)^2$. The accuracy of the angles were estimated from $\delta \theta \sim N_p/ \sqrt{N-1}  \sqrt{\chi ^2_{\textrm{min}}/A}$ where $N_p=3$ is the number of fit parameters, $N=120$ the number of data points, and $A$ the curvature \cite{GreberGd}. 

%
%
To relate the finding of the different angles between the magnetic moments of the Dy ions in Dy$_2$ScN@C$_{80}$ and Dy$_2$TiC@C$_{80}$ with their calculated molecular structures, we first analyzed the distribution of $\mathrm{Dy}-\mathrm{N}-\mathrm{Dy}$ and $\mathrm{Dy}-\mathrm{C}-\mathrm{Dy}$ angles $\theta_{bond}$. The position of the endohedral cluster inside the C$_{80}-I_h$ fullerene cage is flexible and allows different conformes, which may have somewhat different structural parameters. We therefore performed a complete search of possible conformers for Dy$_2$ScN@C$_{80}$ and Dy$_2$TiC@C$_{80}$ by creating 120 different orientations of the cluster inside the fullerene for each molecule using Fibonacci sampling \cite{C9CP00634F} and then performing their DFT optimization. The procedure resulted in 3 and 7 unique conformers for nitride and carbide, respectively, with an energy spread of 47 and 98 meV (Fig. \ref{fig3}, Tabe S1). Figure \ref{fig3} shows their relative energies and $\mathrm{Dy}-\mathrm{X}-\mathrm{Dy}$ angles. Averaging the angles with Boltzmann factors gave $116.2^{\circ}$ and $111.5^{\circ}$ for Dy$_2$ScN@C$_{80}$ and Dy$_2$TiC@C$_{80}$, respectively. However, the orientation of the quantization axes of the Dy ions do not exactly coincide with the $\mathrm{Dy}-\mathrm{N}$ or $\mathrm{Dy}-\mathrm{C}$ bond directions. To consider this aspect, we performed CASSCF/RASSI-SO calculations to determine orientation of quantization axes for each Dy ion and hence the angle between them in each conformer. Calculations showed deviation of the quantization axes from the $\mathrm{Dy}-\mathrm{X}$ bond by $2-5^{\circ}$ for most of the structures. The angles between magnetic moments deviate from the geometrical $\mathrm{Dy}-\mathrm{X}-\mathrm{Dy}$ angles by  $1^{\circ}-2^{\circ}$ for Dy$_2$ScN@C$_{80}$ and $\sim 4^{\circ}$ for Dy$_2$TiC@C$_{80}$ (see S.I. for exact values in each conformer). Weighted with Boltzmann factors, averaged angles between magnetic moments in Dy$_2$ScN@C$_{80}$ and Dy$_2$TiC@C$_{80}$ from \textit{ab initio} calculations are $116.7^{\circ}$ and $115.4^{\circ}$, and are in a remarkably good agreement with the experimental results. It is necessary to keep some reservations since calculations for the conformers are performed for isolated molecules, whereas intermolecular interaction may affect the energies and angle distributions. Yet we get the consistent result that, both the geometrical  $\mathrm{Dy}-\mathrm{X}-\mathrm{Dy}$ angles and the angle between the Dy$^{3+}$ magnetic moments in Dy$_2$TiC@C$_{80}$ is smaller than in Dy$_2$ScN@C$_{80}$. Furthermore, the difference between the geometrical and the magnetic angle is larger for Dy$_2$TiC@C$_{80}$, which indicates a lower axial symmetry.\\

%

The barrier $U_{FA}$ corresponding to the excitation energy between the ferromagnetic and antiferromagnetic doublets  has two components

\begin{equation}
U_{FA} = \Delta E^{\mathrm{dip}}_{FA} + \Delta E_{FA}^{\mathrm{ex}}
\end{equation}

where $\Delta E^{\mathrm{dip}}_{FA}$ and $\Delta E_{FA}^{\mathrm{ex}}$ are the differences in dipolar and exchange energies between the two doublets, respectively. A decreased magnetic bi-stability in didysprosium endofullerene SMMs is typically associated with weaker exchange coupling, as recently demonstrated for Dy$_2$LuN@C$_{80}$ where the barrier is almost entirely due to dipole-dipole interactions \cite{Dy2LuN}. The dipole contribution 

\begin{equation}
\Delta E^{\mathrm{dip}}_{FA}= E_0^{\mathrm{dip}}[3 - \cos(\theta_{mag})]
\label{dip}
\end{equation}

depends on the mutual orientation of the two moments $\theta_{mag}$ and a constant term $E^{\mathrm{dip}}_0=\mu_0 \mu_1 \mu_2/4\pi r^3_{12}$ that only depends on the magnitude of the single-ion magnetic moments $\mu$ and the $\mathrm{Dy}-\mathrm{Dy}$ distance $r_{12}$. The dipole contribution to the barrier can be estimated using the experimentally determined angles $\theta_0$, $r_{12}^{\mathrm{ScN}}=3.58$~\AA~ and $r_{12}^{\mathrm{TiC}}=3.61$~\AA~ as obtained from the DFT optimized geometries, and the nominal values of $10\mu_B$ for the Dy$^{3+}$ moments. The resulting dipole-dipole interaction is $\sim 5 \%$ larger in Dy$_2$ScN@C$_{80}$. Using a Hamiltonian reminiscent of Heisenberg and Lines \cite{Lines,LnSMMS}, the exchange component 

\begin{equation}
\Delta E_{FA}^{\mathrm{exc}} \propto |j_{ex}\cos(\theta_{mag})|
\label{ex}
\end{equation}

is proportional to the strength of the exchange coupling $|j_{ex}|$ and the angle $\theta_{mag}$ between the pseudospins. With the same coupling, $j_{ex}$, the orientation of the moments in Dy$_2$ScN@C$_{80}$ would again be favorable with a $\sim 7\%$ increase over Dy$_2$TiC@C$_{80}$. Thus, the larger exchange and dipole barrier in Dy$_2$TiC@C$_{80}$ can not be explained by the orientation of the moments and the $\mathrm{Dy}-\mathrm{Dy}$ distance, and is therefore the result of stronger exchange interactions.

\begin{figure}[h]
\begin{center}
\includegraphics[width=8.5cm]{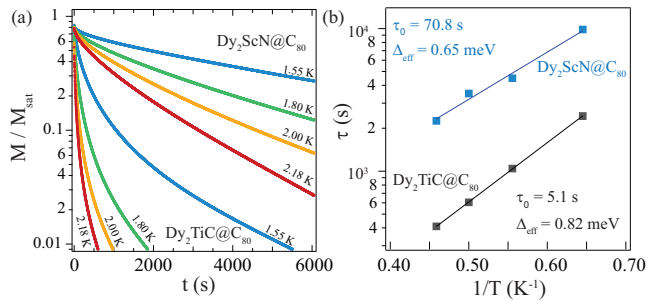}
	\caption{(a)Relaxation of the remanent magnetization  $M$ after saturation $M_{\mathrm{sat}}$ at 7 T.  (b) Arrhenius plot of the extracted relaxation rates, see \cite{SI}}
\label{fig4}
\end{center}
\end{figure}

Insight into the relaxation mechanism is obtained from the Arrhenius plots of the magnetic life times. Figure \ref{fig4} (a)  shows remanent relaxation curves for Dy$_2$ScN@C$_{80}$ and Dy$_2$TiC@C$_{80}$ on a natural logarithm scale after being magnetized at 7 T. From the graph, it is clear that Dy$_2$TiC@C$_{80}$ exhibits a faster relaxation of the magnetization compared to Dy$_2$ScN@C$_{80}$. Figure \ref{fig4} (b) shows the extracted relaxation times (see \cite{SI}) from which kinetic barriers of $\Delta_{\mathrm{eff}}^{\mathrm{ScN}}=0.65$ meV and $\Delta_{\mathrm{eff}}^{\mathrm{TiC}}=0.82$  meV are obtained. The value of 0.65 meV for Dy$_2$ScN@C$_{80}$ is close to that of $\Delta_{\mathrm{eff}}=0.73$ meV  \cite{westerstromPrb14} and 0.77 meV \cite{GreberGd} from previous studies, and again lower than that of Dy$_2$TiC@C$_{80}$. 
 The prefactors or interception points with 1/T=0 in the Arrhenius plot,  $\tau_0$,  get 5.1 and 70.8 s, for  Dy$_2$TiC@C$_{80}$ and Dy$_2$ScN@C$_{80}$, respectively. This is another important indicator for the magnetic stability \cite{GreberGd}, though it is particularly difficult to predict.

We recall that $\Delta_{\mathrm{eff}}$ is a kinetic barrier that is met on the approach of the thermal equilibrium, while $U_{FA}$ is an energy difference as obtained from the equilibrium magnetization curves. Therefore, the two quantities must not be the same, as it was recently shown for the case of Dy$_2$GdN@C$_{80}$ \cite{GreberGd}. Nevertheless, neither a larger exchange and dipole barrier  $U_{FA}$ nor a higher kinetic barrier $\Delta_{\mathrm{eff}}$ in Dy$_2$TiC@C$_{80}$ appears to lead to a better stabilization of the remanent magnetization compared to Dy$_2$ScN@C$_{80}$. Having established that the magnitude of the barriers alone can not explain the decreased magnetic bistability in Dy$_2$TiC@C$_{80}$, we turn to the prefactors $\tau_0$ for the decay process. Previous studies of dinuclear lanthanide endofullerene SMMs have shown that the prefactors can play a more significant role than the height of the barrier in the relaxation process \cite{GreberTb2, GreberGd, Chen2017}. In the present study, the prefactor for Dy$_2$TiC@C$_{80}$ is more than one order of magnitude smaller than for Dy$_2$ScN@C$_{80}$, which  overcompensates the increased barrier and leads to a decreased magnetic bistability. The above analysis of the angles between the Dy magnetic moments, and the comparison to theoretical $\mathrm{Dy}-\mathrm{N}-\mathrm{Dy}$ and $\mathrm{Dy}-\mathrm{C}-\mathrm{Dy}$ bond-angles indicate a lower axial symmetry for Dy$_2$TiC@C$_{80}$, which is in line with a higher relaxation rate.

\section{CONCLUSIONS} 

In conclusion, we have demonstrated that the angle between the magnetic moments on the two Dy sites in the endofullerenes SMMs  DySc$_2$N@C$_{80}$ and Dy$_2$TiC@C$_{80}$ can be determined directly from temperature-dependent magnetization curves to an accuracy better than 1$^{\circ}$. 
Comparing the two systems further reveal a $\sim 49\%$ larger exchange and dipole barrier in Dy$_2$TiC@C$_{80}$, even though the remanent magnetization lifetimes are much shorter. The barrier extracted from Arrhenius plots of the remanent relaxation rates is again larger for Dy$_2$TiC@C$_{80}$. However, the prefactor for the relaxation process is more than one order of magnitude smaller than for DySc$_2$N@C$_{80}$, which overcompensates the increased barrier height and leads to faster relaxation of the remanent magnetization.
The experimentally determined angles between the magnetic moments are in perfect agreement with the theoretical prediction and improve the confidence into the theory that finds a larger deviation between the orientation of the magnetic moments and the bond-angles between the dysprosium ions and the central nitrogen or carbon ion for Dy$_2$TiC@C$_{80}$. This is in line with the observed higher relaxation rate of the magnetisation and a step toward a better quantitative understanding of the bistability in single molecule magnets.

\subsection{Acknowledgements}

This work was financed by the Swedish Research Council (Grant No. 2015-00455), Sklodowska Curie Actions co-founding project INCA 600398 and profited from the Swiss National Science Foundation grant No 206021-150784. A.A.P and  and S. M. Aacknowledges Deutsche Forschungsgemeinschaft for financial support (grants PO 1602/5, PO 1602/7, and AV 169/3).

\bibliography{REF_2p3}

\end{document}